\documentclass[11pt]{article}
\usepackage{amsmath,amssymb}%
\usepackage{pdflscape}
\usepackage{hyperref}
\usepackage{color}
\usepackage{verbatim}

\usepackage{graphicx}  

\usepackage{color}

\usepackage{psfrag}
\usepackage{epsfig}
\usepackage{epstopdf}
\usepackage{amssymb,latexsym}
\usepackage[mathscr]{eucal}
\usepackage{cite}
\usepackage[active]{srcltx}
\usepackage{url}
\usepackage{ntheorem}

\DeclareFontFamily{OT1}{rsfs}{} \DeclareFontShape{OT1}{rsfs}{m}{n}{
<-7> rsfs5 <7-10> rsfs7 <10-> rsfs10}{}
\DeclareMathAlphabet{\mycal}{OT1}{rsfs}{m}{n}

{\catcode `\@=11 \global\let\AddToReset=\@addtoreset}
\AddToReset{equation}{section}

{\catcode `\@=11 \global\let\AddToReset=\@addtoreset}
\AddToReset{figure}{section}

{\catcode `\@=11 \global\let\AddToReset=\@addtoreset}
\AddToReset{table}{section}

\newcounter{mnotecount}[section]

\newcommand{\ednote}[1]{}

\definecolor{bluem}{rgb}{0,0,0.5}

\definecolor{mycolor}{cmyk}{0.5,0.1,0.5,0}
\definecolor{michel}{rgb}{0.5,0.9,0.9}

\definecolor{turquoise}{rgb}{0.25,0.8,0.7}
\definecolor{bluem}{rgb}{0,0,0.5}

\definecolor{MDB}{rgb}{0,0.08,0.45}
\definecolor{MyDarkBlue}{rgb}{0,0.08,0.45}

\definecolor{MLM}{cmyk}{0.1,0.8,0,0.1}
\definecolor{MyLightMagenta}{cmyk}{0.1,0.8,0,0.1}

\definecolor{HP}{rgb}{1,0.09,0.58}
\definecolor{dkgreen}{rgb}{0,0.8,0}

\newcommand{\dt}{\partial}
\newcommand{\R}{\mathbb{R}}

\begin{document}
\title{Degenerating Black Saturns}

\author{Micha\l\  Eckstein\thanks{Faculty of Mathematics and Computer Science, Jagellonian University, michal.eckstein@uj.edu.pl}}
\date{}
\maketitle{}
\begin{abstract}
We investigate the possibility of constructing degenerate Black Sa-turns in the family of solutions of Elvang-Figueras. We demonstrate that such solutions suffer from naked singularities.
\end{abstract}


\section{Introduction}

In \cite{EF} Elvang and Figueras have presented a family of axisymmetric black hole solutions to vacuum 4+1-dimensional Einstein equations. Due to the specific topology of the event horizon: $\R\times\big( (S^1\times S^2) \cup S^3)\big)$ it has been named Black Saturn. It can be regarded as a spherical Myers-Perry black hole \cite{MP} surrounded by a black ring \cite{ER,PS}. The configuration is kept in balance by the angular momenta.

The Black Saturn metrics are of great significance since they provide an example of well-behaved stationary black hole space-times with disconnected Killing horizon. This shows a sharp contrast between solutions to Einstein equations in 4+1 and 3+1-dimensions since, as proven recently \cite{NH1,NH2,NH3,N2K}, in the latter case analytic stationary two black hole space-times are nakedly singular.

The family of Black Saturn solutions is constructed via the inverse scattering method \cite{BZ} which introduces 8 real parameters: $a_i$ with $ i = 1, \ldots , 5$ and $c_1, \, c_2, \, k$. Moreover, a ninth one - $q$ is brought in by a change of coordinates to facilitate the asymptotic flatness (see \cite[p. 10, footnote 3]{EF}). The ordering of parameters $a_i$ assumed in \cite{EF} reads
\begin{align} \label{aOrd}
 a_1\le a_5 \le a_4 \le a_3 \le a_2
\end{align}
or in terms of the dimensionless ones $\kappa_i = \frac{a_{i+2}-a_1}{(a_2-a_1)^2}$,
\begin{align*}
0 \leq \kappa_3 \leq \kappa_2 \leq \kappa_1 \leq 1.
\end{align*}

A thorough analysis (see \cite{EF,MY,Sebastian}) has
 shown that, under the assumption of parameters $a_i$ being pairwise distinct,
  the metrics in the Black Saturn family describe asymptotically flat, stably causal black hole space-times with smooth domains of outer communications. To guarantee the above listed desired properties of a well-behaved black hole space-time one needs to tune the parameters $c_1, \, c_2, \, k$ and $q$ in terms of $a_i$'s.

The purpose of this work is to investigate the possibility of obtaining a well-behaved metric from the Black Saturn family in the case of coalescence of some of the parameters $a_i$. Such coalescence corresponds to the ``pole fusion effect''
  in the inverse scattering method, which may lead to extremal black-hole solutions (see \cite[Chapter 8.3]{BZ}). There are however various ways (paths in the parameter space) in which one can obtain a coalescence of two or more $a_i$'s. The result, will \textit{a priori} depend on the chosen limiting procedure as demonstrated in \cite{Geroch} \footnote{We thank Sebastian Szybka for pointing out this reference to us.}. For instance, in \cite[Section A.1]{EF} it has been shown that to obtain the limiting case of a Myers-Perry black hole from the balanced Black Saturn configuration one needs to take first $a_5 \nearrow a_4$ and then $a_1 \nearrow a_4$. In general, the assumption that the coalescence is to be considered after the fine tuning of parameters $c_1, \, c_2, \, k, \, q$ already imposes restrictions on the limiting procedure, since, for example, $a_1 \nearrow a_5$ causes $c_1$ to diverge (see \cite[(3.7)]{EF} or \cite[(2.3)]{MY}). Let us note that the parameters $c_1$ and $c_2$ may \textit{a priori} assume infinite values. Indeed, the line element \eqref{SaturnMetric} has a well-defined limit for $c_1 \to \pm\infty$ and/or $c_2 \to \pm\infty$, which moreover commutes with  every coalescence considered in this paper. However, these cases need separate analysis of possible balance conditions.
  
When the balance conditions are imposed on the Black Saturn solution, the areas of the horizons of the two disconnected components \cite[(3.26, 3.27)]{EF} tend to zero in the limits $a_3 \nearrow a_2$ and $a_5 \nearrow a_4$ respectively. This suggests that the possible degenerate solutions are nakedly-singular. However, there is no \textit{a priori} reason for the procedures of coalescence of parameters and imposition of the balance conditions to commute, so this observation does not exclude the possibility of obtaining well-behaved extremal solutions via some other limiting procedure. Let us note, that the same coalescence that leads to the vanishing of the horizon area of the black ring component implies the divergence of its temperature \cite[(3.28)]{EF}. This suggests that the limiting procedures adopted in \cite{EF} are not the right ones, as one should expect $T = 0$ for an extremal black-hole solution, since the temperature is proportional to surface gravity.

The strategy we adopt in this paper is to consider the limits $a_i \to a_j$ at the level of metric functions of the full Black Saturn solution and then investigate whether the balance conditions can be fulfilled by a fine tuning of parameters $c_1, \, c_2, \, k$ and $q$. To make the paper self-contained we present in the Appendix the Black Saturn metric of \cite{EF} in generalised Weyl coordinates. For the details of construction and properties we refer the reader to \cite{EF} and \cite{MY,Sebastian}.

\section{Analysis}

Since we are interested only in the solutions with two disconnected components of the event horizon (compare with the rod structure \cite[Figure 1]{EF}), we shall assume the strict inequality $a_4 < a_3$ in the ordering \eqref{aOrd}. We have thus 3 possible two-fold coalescences to be considered in the next subsections. Moreover, there are 3 three-fold and 1 four-fold limit that need to be investigated. When more then two $a_i$ parameters coalesce, one can consider various different paths in the parameter space that lead to the same coalescence. Fortunately, if the limiting procedure is performed at the level of the metric, the ordering of the limits does not play a role (compare \cite[Section A.1]{EF}). This is because $a_i \nearrow a_j$ implies $\mu_i \nearrow \mu_j$ \eqref{mu} and all of the metric functions (see Appendix) are smooth as functions of $\mu_i$'s.

In each of the subsections we consider a particular coalescence of the $a_i$'s parameters while keeping the other distinct. The reason for that is that the behaviour of the metric functions on the axis ($\rho = 0$) should be studied separately in each region of the axis $a_i \leq z \leq a_j$ (see \cite[Section 5.4]{MY}). This means that each coalescence needs a separate procedure of investigation of the metric functions on the axis.

The detailed analysis of the regularity, asymptotic flatness and causality of the seven limiting cases of the Black Saturn solution is straightforward, but lengthy - one essentially follows the strategy adopted in \cite{MY}. However, since our analysis shows that in neither of the investigated limits can one tune the parameters to obtain a balanced configuration we shall only present the part of reasoning that leads to this conclusion.

\subsection{\texorpdfstring{$a_1 \nearrow a_5$}{a1 --> a5}}\label{a1a5}

Let us note first, that if one takes the limit $a_1 \nearrow a_5$, then the resulting metric does not depend on the parameter $c_1$ anymore. Indeed, $\mu_1 = \mu_5$ implies $M_1 = M_3 = M_4 = 0$ (see Appendix), thus the parameter $c_1$ completely drops out of the line element. According to \cite[p. 7]{EF} this configuration would describe a static black ring around an $S^3$ black hole, which are kept apart by a conically singular membrane. Indeed, one can detect the conical singularity by investigating the periodicity of the variable $\varphi$ (compare \cite[Section 4]{MY}). To avoid conical singularity at zeros of the Killing vector $\dt_{\varphi}$ one needs the ratio
\begin{align*}
\lim_{\rho \to 0} \, \frac{\rho^2 g_{\rho \rho}}{g_{\varphi \varphi}}
\end{align*}
to be constant on the set $\{z < a_1\} \cup \{a_4 < z < a_3\}$, which is an axis of rotation for $\dt_{\varphi}$. By investigating the leading behaviour in $\rho$ of the metric functions $g_{\varphi \varphi}$ and $g_{\rho \rho}$ in the relevant region of the space-time we obtain
\begin{align*}
\lim_{\rho \to 0} \, \frac{\rho^2 g_{\rho \rho}}{g_{\varphi \varphi}} =
	\begin{cases}
		k^2, & \text{for } z < a_1 \\
		k^2 \, \frac{ (a_2-a_1) (a_3-a_4)^2}{(a_3-a_1)^2 (a_2-a_4)}, & \text{for } a_4 < z < a_3
	\end{cases}.
\end{align*}
Hence, to avoid conical singularities one would need to have
\begin{align*}
\frac{ (a_2-a_1) (a_4-a_3)^2}{(a_3-a_1)^2 (a_2-a_4)} = 1,
\end{align*}
which is equivalent to
\begin{align*}
a_4 = a_1 && \text{ or } && a_4 = \frac{a_1 a_2 - 2 a_2 a_3 + a_3^2}{a_1 - a_2}.
\end{align*}
The first case is excluded, whereas the second one would require
\begin{align*}
\frac{a_1 a_2 - 2 a_2 a_3 + a_3^2}{a_1 - a_2} < a_3,
\end{align*}
as $a_4 < a_3$ by assumption. The latter however imply that either $a_3 < a_1$ or $a_3 > a_2$, which contradicts the ordering \eqref{aOrd}.

This means that the conical singularity on the axis cannot be avoided. 

\subsection{\texorpdfstring{$a_5 \nearrow a_4$}{a5 --> a4}}\label{a5a4}

Let us now investigate the coalescence $a_5 \nearrow a_{4}$. We shall start with the ana-lysis of the Killing vector field $\partial_t$ on the set $\{\rho = 0,\, z  \leq a_1\}$.
A {\sc Mathema-tica} calculation shows that
$g_{tt}$ is a rational function with the denominator given by
\begin{align*}
&  \left(2 (a_3-a_1) (a_2-a_4)+(a_4-a_1) c_1 c_2\right)^2 (z-a_1) (z-a_2) (z-a_4)
      \;,
\end{align*}
which vanishes as $z$ approaches $a_1$ from below. On
the other hand, its numerator has the following limit as $z \nearrow a_1$,
\begin{align*}
(a_2-a_1)^2 (a_3-a_1) (a_4-a_1)^2 \left(2 (a_3-a_1) - c_1^2\right) c_2^2.
\end{align*}
Hence, we have now two possibilities of tuning the parameters to avoid a naked singularity at $\rho=0,z=a_1$:
\begin{flalign}
1. \quad & c_1 = \pm \sqrt{2(a_3-a_1)}, \label{cond2} \\
2. \quad & c_2 = 0. \label{cond3}
\end{flalign}

Keeping them in mind, we shall investigate the behaviour of the Killing vector field $\partial_t$ on the set $\{\rho = 0,\, a_4 \leq z \leq a_3\}$. The function $g_{tt}$ on this domain is a rational function with the denominator
\begin{align*}
2 (a_1 - a_2)^2 (z - a_1) (z - a_2) (z-a_4),
\end{align*}
vanishing at $z=a_4$. On the other hand, the numerator of $g_{tt}$ at $\rho=0$, $z = a_4$ reads
\begin{align*}
(a_1 - a_4)^2 (a_2 - a_4)^2 (c_1 - c_2)^2.
\end{align*}
Thus, there is only one possibility to avoid a naked singularity at $z=a_4$: set $c_1 = c_2$. Combining the results obtained so far we end up with the following possible fine tunings:
\begin{flalign}
1. \quad & c_1 = c_2 = \pm \sqrt{2(a_3-a_1)}, \label{condB1} \\
2. \quad & c_1 = c_2 = 0. \label{condB2}
\end{flalign}

The choice $c_1 = c_2 = 0$ would bring us back to the seed solution \cite{EF}, which is nakedly singular, so we are forced to set $c_1 = c_2 = \pm \sqrt{2(a_3-a_1)}$.

Let us now analyse the behaviour of the Killing vector field $\partial_{\psi}$ on the set $\{\rho = 0,\, a_1 \leq z \leq a_4\}$.
A {\sc Mathematica} calculation shows that
$g_{\psi\psi}$ is a rational function with the denominator given by
\begin{align*}
&  -2 \left((a_2 - a_4) c_1 + (a_4-a_1) c_2\right)^2 (z-a_1) (z-a_2) (z-a_4)
      \;.
\end{align*}
The singularity at $z = a_1$ is cancelled by the tuning \eqref{condB1} since the numerator of $g_{\psi\psi}$ at $z = a_1$ reads
\begin{align*}
&  -(a_2 - a_1)^2 (a_4 - a_1)^2 (2 (a_3 - a_1) - c_1^2) (2 (a_2 - a_4) + c_2 q)^2
      \;.
\end{align*}
On
the other hand, the denominator of $g_{\psi\psi}$ is singular at $z = a_4$ and the numerator has the following limit for $z \nearrow
a_4$,
\begin{multline*}
2 (a_4 - a_1)^2 (a_2 - a_4)^2 (a_3 - a_4) (2 (a_2 - a_1) - (c_1 - c_2) q)^2 =\\
= 8 (a_4 - a_1)^2 (a_2 - a_4)^2 (a_3 - a_4) (a_2 - a_1)^2,
\end{multline*}
which does not vanish. This means that the naked singularity at $\rho = 0$, $z = a_4$ persists regardless of the fine tuning of parameters.

We have so far dealt with the situation of the parameters $c_1$ and $c_2$ assuming finite values. Let us now turn to the case $c_1 \to \pm \infty$. In this instance $g_{tt}$, being the norm of the Killing vector $\dt_t$, is given in the region $\{\rho = 0,\, z  \leq a_1\}$ by the following formula,
\begin{align*}
-\frac{(a_2-z)(a_3-z)}{(a_1-z)(a_4-z)}.
\end{align*}
This expression diverges as $z \nearrow a_1$ and the singularity cannot be cancelled by any fine-tuning of the free parameters.

For $c_2 \to \pm \infty$ we obtain that $g_{tt}$ on the set $\{\rho = 0,\, a_1 \leq z \leq a_4\}$ is a rational function with the denominator,
\begin{align*}
2 (a_4 - a_1)^2 (a_2 - z) (a_4 - z) (z-a_1),
\end{align*}
vanishing at $z = a_4$. On the other hand, its numerator has the following limit for $z \nearrow a_4$,
\begin{align*}
2 (a_4 - a_1)^2 (a_2 - a_4)^2 (a_3 - a_4).
\end{align*}
We conclude, that in this configuration there is a naked singularity at $\rho = 0$, $z = a_4$ that cannot be avoided.

For $c_1, c_2 \to \pm \infty$ we have $g_{tt} = - \frac{\mu_2 \mu_3}{\mu_1 \mu_4}$, which is singular on the axis $\{\rho = 0\}$ in the region $a_1 \leq z \leq a_2$.

\subsection{\texorpdfstring{$a_3 \nearrow a_2$}{a3 --> a2}}\label{a2a3}

Let us now consider the coalescence $a_3 \nearrow a_2$.

To rule out smooth non-trivial solutions it is sufficient to investigate the behaviour of the Killing vector field $\partial_t$ in the region $\{\rho = 0,\, a_4 \leq z \leq a_2\}$. With the help of {\sc Mathematica} we obtain that
$g_{tt}$ is a rational function with the denominator given by
\begin{align*}
&  2 (a_2-a_1)^2 (z-a_1) (a_2-z) (a_5-z),
\end{align*}
which vanishes as $z$ approaches $a_2$ from below. On
the other hand, its numerator has the following limit as $z \nearrow
a_2$,
\begin{align*}
- (a_2 - a_1)^2 (a_2 - a_5)^2 c_2^2.
\end{align*}
This means, that one should impose the condition $c_2 = 0$ to avoid a naked singularity at $z=a_2$. But setting $a_3 = a_2$ and $c_2 = 0$ completely removes the $S^3$ black hole component \cite[Section A.2]{EF} and we are left with a single $\psi$-spinning black ring.

In the case $c_1 \to \pm \infty$ the function $g_{tt}$ in the region $\{\rho = 0,\, z  \leq a_1\}$ reads,
\begin{align*}
-\frac{(a_2-z)^2}{(a_1-z)(a_4-z)}.
\end{align*}
Thus, a naked singularity pops out at $z = a_1$.

For $c_2 \to \pm \infty$, $g_{tt}$ in the region $\{\rho = 0,\, a_5 \leq z \leq a_4\}$ turns out to also be given by,
\begin{align*}
-\frac{(a_2-z)^2}{(a_1-z)(a_4-z)},
\end{align*}
now leading to a singularity at $z = a_4$.

Similarly to the case described in Section \ref{a5a4}, for $c_1, c_2 \to \pm \infty$ we have $g_{tt} = - \frac{\mu_2^2}{\mu_1 \mu_4}$, which becomes singular on the axis $\{\rho = 0\}$ in the whole region $a_1 \leq z \leq a_2$.

\subsection{\texorpdfstring{$a_1 \nearrow a_5 \nearrow a_4$}{a1 --> a5 --> a4}}

According to \cite[Section A.1]{EF} in this limit the Black Saturn metric reduces to a Myers-Perry black hole with a single angular momentum, hence no further analysis is needed. Let us stress however, that to obtain this result independently of the order of the limits one needs to compute the limits at the level of the metric functions - before the imposition of the balance conditions.

\subsection{\texorpdfstring{$a_1 \nearrow a_5, \; a_3 \nearrow a_2$}{a1 --> a5, a3 --> a2}}

Let us first investigate the behaviour of the Killing vector field $\partial_t$ on the set $\{\rho = 0, \, a_4 \leq z \leq a_2\}$. Again with the help of {\sc Mathematica} we obtain the following formula for $g_{tt}$ function in this region
\begin{align*}
\frac{z-a_4}{a_1 - z} + \frac{c_2^2}{2 (a_2 - z)}.
\end{align*}
We have a naked singularity at $z=a_2$ unless we set $c_2 = 0$. As argued in Section \ref{a2a3} this completely removes the $S^3$ black hole component. What is more, the conical singularity detected in Section \ref{a1a5} persists. Indeed, we have
\begin{align*}
\lim_{\rho \to 0} \, \frac{\rho^2 g_{\rho \rho}}{g_{\varphi \varphi}} =
	\begin{cases}
		k^2, & \text{for } z < a_1 \\
		k^2 \, \frac{ a_2-a_4}{a_2-a_1}, & \text{for } a_4 < z < a_2
	\end{cases}.
\end{align*}
Hence, to guarantee the correct periodicity of $\varphi$ we would have to set $a_4 = a_1$, which is excluded by the assumptions of this section.

Since the parameter $c_1$ has dropped out of the line element in the coalescence considered in this Subsection, we need only to comment on the instance $c_2 \to \pm \infty$. In this case, the function $g_{tt}$ in the region $\{z  \leq a_1\}$ behaves near the axis $\{\rho = 0\}$ like
\begin{align*}
\frac{4 (a_1 - z)(a_2 - z)^2}{(a_4 - z) \rho^2} + \mathcal{O}(\rho^0).
\end{align*}
This excludes the possibility of $c_2 \to \pm \infty$ leading to a well-behaved space-time.


\subsection{\texorpdfstring{$a_5 \nearrow a_4, \; a_3 \nearrow a_2$}{a5 --> a4, a3 --> a2}}

It is sufficient to analyse the behaviour of the Killing vector field $\partial_t$ on the axis. In the region $\{\rho = 0,\, z \leq a_1\}$ 
$g_{tt}$ is a rational function with the denominator given by
\begin{align*}
&  \left(2 (a_2-a_1) (a_2-a_4)+(a_4-a_1) c_1 c_2\right)^2 (z-a_1) (z-a_2) (z-a_4)
      \;.
\end{align*}
As $z \nearrow a_1$ its numerator reads
\begin{align*}
(a_2-a_1)^3 (a_4-a_1)^2 \left(2 (a_2-a_1) - c_1^2\right) c_2^2.
\end{align*}
Thus, to avoid a naked singularity at $\rho=0,z=a_1$ one has to set 
\begin{align} \label{c1c2}
c_1 = \pm \sqrt{2(a_2-a_1)}, && \text{or} && c_2 = 0.
\end{align}

Let us now switch to the region $\{\rho = 0,\, a_1 \leq z  \leq a_4\}$. A {\sc Mathematica} calculation shows that
$g_{tt}$ is a rational function with the denominator equal to
\begin{align*}
&  2 \left( (a_2 - a_4) c_1 + (a_4 - a_1) c_2 \right)^2 (z-a_1) (z-a_2) (z-a_4)
      \;.
\end{align*}
The continuity of $g_{tt}$ at $z=a_1$ is easily verified for both choices of parameters \eqref{c1c2}. On the other hand, as $z$ approaches $a_4$, $g_{tt}$ becomes singular since its numerator at $z=a_4$ reads
\begin{align*}
2 (a_1 - a_4)^2 (a_2 - a_4)^3 (c_1 - c_2)^2.
\end{align*}
To bypass the naked singularity at $\rho=0,z=a_4$ we need to set $c_1 = c_2$ in addition to \eqref{c1c2}.

Finally, in the region $\{\rho = 0,\, a_4 \leq z  \leq a_2\}$ the denominator of $g_{tt}$ is given by
\begin{align*}
2 (a_1 - a_2)^2 (z-a_1) (z-a_2) (z-a_4).
\end{align*}
Again, the continuity of $g_{tt}$ at $z=a_4$ is guaranteed by the tuning of para-meters imposed so far. However, the numerator of $g_{tt}$ at $z=a_2$ reads
\begin{align*}
-(a_1 - a_2)^2 (a_2 - a_4)^2 c_2^2,
\end{align*}
so the only way to avoid a singularity at $z=a_2$ is to set $c_2=0$. Combining this with the previous results we conclude that to assure the smoothness of the Killing vector field $\partial_t$ on the axis $\{\rho =0\}$ one needs to set $c_1 = c_2 = 0$. As already argued, this would bring us back to the seed solution \cite{EF}, which is singular itself.

It remains to check the possibility of cancelling the singularities by letting one or both of the parameters $c_1, c_2$ go to $\pm \infty$. As $c_1 \to \pm \infty$ we obtain that $g_{tt}$ in the region $\{\rho = 0, \, z  \leq a_1\}$ is given by the expression
\begin{align*}
-\frac{(a_2-z)^2}{(a_1-z)(a_4-z)},
\end{align*}
singular at $z=a_1$.

For $c_2 \to \pm \infty$ on the other hand, we obtain the following behaviour of $g_{tt}$ near the axis $\{\rho = 0\}$ in the region $\{a_4 \leq z  \leq a_2\}$,
\begin{align*}
\frac{4 (a_2 - z)^2 (a_4-z)}{(a_1 - z) \rho^2} + \mathcal{O}(\rho^0).
\end{align*}

Moreover, if we let both $c_1$ and $c_2$ tend to infinity we again obtain $g_{tt} = - \frac{\mu_2^2}{\mu_1 \mu_4}$.

We conclude that the Black Saturn solution with $a_5 \nearrow a_4, \; a_3 \nearrow a_2$ and one or both of the $c_i$ parameters infinite is nakedly singular.

\subsection{\texorpdfstring{$a_1 \nearrow a_5 \nearrow a_4, \; a_3 \nearrow a_2$}{a1 --> a5 --> a4, a3 --> a2}}

As in the previous cases (see Section \ref{a1a5}) the limit $a_1 \nearrow a_5$ implies that the parameter $c_1$ is no longer present in the line element. Furthermore, an investigation of the behaviour of the Killing vector $\dt_t$ on the axis forces us to impose $c_2 = 0$. Indeed, in the region $\{\rho = 0, \, a_1 \leq z \leq a_2\}$ the metric function $g_{tt}$ reads
\begin{align*}
\frac{2(z - a_2) + c_2^2}{2 (a_2 - z)},
\end{align*}
so only $c_2=0$ allows to avoid a singularity at $z=a_2$. But if $c_1$ drops out of the metric functions and $c_2$ vanishes we are again back at the seed solution \cite{EF}, which is of no physical interest.

Moreover, in the case $c_2 \to \pm \infty$ we obtain $g_{tt} = \frac{\mu_2^2}{\rho^2}$, that clearly leads to singularities on the axis.

\section{Conclusions}

We have investigated various different coalescences of parameters defining the Black Saturn solution. We have shown that either the resulting metric is nakedly singular or it reduces to a black hole with one connected component of the event horizon: a Myers-Perry black hole or Emperano-Reall black ring.

Led by the example given by Geroch in \cite{Geroch} one might think that there can still be a way of obtaining a meaningful coalescence limit in the Black Saturn family by employing a smart change of coordinate chart. However, as demonstrated in \cite{Geroch}, the Killing vectors are inherited by any limit of a space-time with some parameters. Strictly speaking, this property has been demonstrated for a 3+1-dimensional case. Nevertheless, as the technique developed in \cite[Appendix B]{Geroch} is general, the proof can be adapted in a straightforward way to a 4+1 dimensional space-time with three Killing vectors. Now, since our analysis consisted in uncovering singularities in the norms of Killing vector fields, we conclude that any coordinate transformation would either lead to the same results or not yield a proper limit space-time at all.

We have thus exhausted the possibility of constructing a smooth extremal Black Saturn configuration in the family of solutions of Elvang-Figueras.

This outcome is in consent with the known properties of $4+1$-dimensional black holes. Both spherical black holes \cite{MP} and black rings \cite{ChruscielCortier,extremal} require two non-vanishing angular momenta to admit smooth extremal configurations. Unfortunately, the Black Saturn solution of Elvang-Figueras has angular momentum in a single plane only and it is not clear if doubly-spinning components can at all be kept in balance \cite{EF}. Thus, the question of existence of smooth stationary axisymmetric black hole with disconnected degenerate Killing horizons in $4+1$ dimensions remains open.

\section*{Acknowledgements}
We would like to thank Piotr T. Chru\'sciel for suggesting this problem to us and for his illuminating remarks. We also thank Patryk Mach and Sebastian Szybka for comments on the manuscript.

The main part of the calculations was carried out using {\sc Mathematica} 8.0.4 by Wolfram Research.

Project operated within the Foundation for Polish Science IPP Programme ``Geometry and Topology in Physical Models'' co-financed by the EU European Regional Development Fund, Operational Program Innovative Economy 2007-2013. Partial support of the Polish Ministry of Science and Higher Education under the grant for young scientists and PhD students is acknowledged.

\appendix

\section{Appendix}

In the generalised Weyl coordinates $(t,\rho,z,\psi,\varphi)$ the black saturn line element~\cite{EF} reads
\begin{multline}
  ds^2 = g_{tt} dt^2 + g_{t \psi} dt d\psi + g_{\psi \psi} d\psi^2 + g_{\rho \rho} d\rho^2 + g_{zz} dz^2 + g_{\varphi \varphi} d\varphi^2 \\
 =
  -\frac{H_y}{H_x} \Big[dt + \Big(\frac{\omega_\psi}{H_y}+q\Big) \,
  d\psi \Big]^2
  + H_x \bigg\{ k^2 \, P \Big( d\rho^2 + dz^2 \Big)
       + \frac{G_y}{H_y} \, d\psi^2 + \frac{G_x}{H_x}\, d\varphi^2 \bigg\} \, ,
 \label{SaturnMetric}
\end{multline}
where $k$, $q$ are real constants. The metric functions depend only on variables $\rho$ and $z$. Define
\begin{align}\label{mu}
 \mu_i :=\sqrt{\rho^2
 + (z-a_i)^2}- (z-a_i)
  \;,
\end{align}
where the $a_i$'s are real constants. The assumed ordering of $a_i$'s \eqref{aOrd} implies
\begin{align*}
\mu_1 \leq \mu_5 \leq \mu_4 \leq \mu_3 \leq \mu_2 && \text{and} && \mu_i = \mu_j \; \Leftrightarrow \; a_i = a_j.
\end{align*}

Let us list the functions constituting the line element \ref{SaturnMetric}:
\begin{align*}
  G_x = \frac{\rho^2\mu_4}{\mu_3\, \mu_5} \, ,
\end{align*}
\begin{align*}
  P =  (\mu_3\, \mu_4+ \rho^2)^2
      (\mu_1\, \mu_5+ \rho^2)
      (\mu_4\, \mu_5+ \rho^2) \, ,
\end{align*}
\begin{align*}
   & H_x = F^{-1} \,
   \bigg[ M_0 + c_1^2 \, M_1 + c_2^2\,  M_2
   +  c_1\, c_2\, M_3 + c_1^2 c_2^2\, M_4 \bigg] \, , 
   \\
   & H_y = F^{-1} \,
   \frac{\mu_3}{\mu_4}\,
   \bigg[ M_0 \frac{\mu_1}{\mu_2}
   - c_1^2 \, M_1 \frac{\rho^2}{\mu_1\,\mu_2}
   - c_2^2\,  M_2 \frac{\mu_1\,\mu_2}{\rho^2}
   +  c_1\, c_2\, M_3
   + c_1^2 c_2^2\, M_4 \frac{\mu_2}{\mu_1} \bigg] \, ,\\&&
\end{align*}
where $c_1$ and $c_2$ are real constants, and
\begin{align*}
  & M_0 = \mu_2\, \mu_5^2 (\mu_1-\mu_3)^2 (\mu_2-\mu_4)^2
   (\rho^2+\mu_1\,\mu_2)^2(\rho^2+\mu_1\,\mu_4)^2
   (\rho^2+\mu_2\,\mu_3)^2 \, ,\\&& \\[2mm]
\nonumber & M_1 = \mu_1^2 \, \mu_2 \, \mu_3\, \mu_4 \, \mu_5 \,
\rho^2\,
  (\mu_1-\mu_2)^2 (\mu_2-\mu_4)^2(\mu_1-\mu_5)^2
  (\rho^2+\mu_2\,\mu_3)^2  \, ,\\&& \\[2mm]
\nonumber
  & M_2 = \mu_2 \, \mu_3\, \mu_4 \, \mu_5 \, \rho^2\,
  (\mu_1-\mu_2)^2 (\mu_1-\mu_3)^2
  (\rho^2+\mu_1\,\mu_4)^2(\rho^2+\mu_2\, \mu_5)^2  \, ,\\&& \\[2mm]
  & M_3 = 2 \mu_1 \mu_2 \, \mu_3\, \mu_4 \, \mu_5 \,
  (\mu_1-\mu_3) (\mu_1-\mu_5)(\mu_2-\mu_4)
  (\rho^2+\mu_1^2)(\rho^2+\mu_2^2) \nonumber \\[1mm]
  &\hspace{3cm} \times
  (\rho^2+\mu_1\,\mu_4)(\rho^2+\mu_2\, \mu_3)
  (\rho^2+\mu_2\, \mu_5)  \, ,\\[2mm]
  & M_4 = \mu_1^2 \, \mu_2\, \mu_3^2 \, \mu_4^2 \,
  (\mu_1-\mu_5)^2
  (\rho^2+\mu_1\,\mu_2)^2(\rho^2+\mu_2\, \mu_5)^2  \, ,
\end{align*} and \begin{multline*}
  F = \mu_1\, \mu_5\,  (\mu_1-\mu_3)^2(\mu_2-\mu_4)^2
  (\rho^2+\mu_1\,\mu_3)
  (\rho^2+\mu_2\,\mu_3)
  (\rho^2+\mu_1\,\mu_4) \nonumber\\
   \hspace{1cm} \times
  (\rho^2+\mu_2\,\mu_4)
  (\rho^2+\mu_2\,\mu_5)
  (\rho^2+\mu_3\,\mu_5)
  \prod_{i=1}^5 (\rho^2+\mu_i^2) \, .
\end{multline*}
Furthermore,
\begin{align*}
  G_y = \frac{\mu_3\, \mu_5}{\mu_4} \, ,
\end{align*}
and the off-diagonal part of the metric is governed by
\begin{align*}
  \omega_\psi
  =
  2 \frac{
     c_1\, R_1\, \sqrt{M_0 M_1}
    -c_2\, R_2\, \sqrt{M_0 M_2}
    +c_1^2\,c_2\, R_2\, \sqrt{M_1 M_4}
    -c_1\,c_2^2\, R_1\, \sqrt{M_2 M_4}
  }
  {F \sqrt{G_x}} \, ,
 \nonumber
\end{align*}
with $R_i = \sqrt{\rho^2 + (z-a_i)^2}$.

The determinant of the metric reads
\begin{align*}
 \det g_{\mu_\nu} = -\rho^2 H_x^2 k^4 P^2
 \;.
\end{align*}

\end{document}